\documentclass[final,3p,times,twocolumn]{elsarticle}

\usepackage{enumerate}
\usepackage{amsmath}
\usepackage{amsfonts}
\usepackage{amssymb}
\usepackage{sectsty}
\usepackage{graphicx} 
\usepackage[english]{babel}
\usepackage{array}
\usepackage[T1]{fontenc}
\usepackage{setspace}
\usepackage{cleveref}
\usepackage[normalem]{ulem}
\usepackage{comment}
\usepackage{soul}
\usepackage{xcolor}
\usepackage{chemfig}
\biboptions{sort&compress}

\journal{Journal of Colloid and Interface Science}

\begin{document}

\begin{frontmatter}

\title{Simple ions control the elasticity of calcite  gels via interparticle forces}

\author[ILM,TUWien]{Teresa Liberto}
\author[ILM,IUF]{Catherine Barentin\corref{mycorrespondingauthor}}
\cortext[mycorrespondingauthor]{Corresponding author}
\ead{catherine.barentin@univ-lyon1.fr}
\author[ILM]{Jean Colombani}
\author[ISTEC]{{Anna Costa}}
\author[ISTEC]{{Davide Gardini}}
\author[Milano]{{Maurizio Bellotto}}
\author[ILM]{Marie Le Merrer}

\address[ILM]{Universit\'e de Lyon, Universit\'e Claude Bernard Lyon 1, CNRS, Institut Lumi\`ere Mati\`ere, F-69622, VILLEURBANNE, France}
\address[TUWien]{Current address: Faculty of Civil Engineering, Vienna University of Technology, Adolf Blamauergasse 1-3, A-1030 Vienna, Austria.}
\address[IUF]{Institut Universitaire de France}
\address[ISTEC]{CNR-ISTEC, Institute of Science and Technology for Ceramics - National Research Council of Italy, Via Granarolo 64, I-48018 Faenza, RA, Italy}
\address[Milano]{Dipartimento di Chimica, Materiali ed Ingegneria Chimica "G. Natta", Politecnico di Milano, Piazza Leonardo da Vinci 32, 20133 Milano, Italy}

\begin{keyword}
Colloidal gel \sep Calcite \sep DLVO \sep Zeta potential 
\end{keyword}

\begin{abstract}

Suspensions of calcite in water are employed in many industrial fields such as paper filling, pharmaceutics, heritage conservation or building construction, where the rheological properties of the paste need to be controlled.
We measure the impact of simple ions such as calcium, sodium or hydroxide on the elasticity of a nanocalcite paste, which behaves as a colloidal gel.
We confront our macroscopic measurements to DLVO interaction potentials, based on chemical speciations and measurements of the zeta potential.
By changing the ion type and concentration, we go beyond the small repulsion regime  and span two orders of magnitude in shear modulus.
Upon addition of calcium hydroxide, we observe a minimum in shear modulus, correlated to a maximum in the DLVO energy barrier, due to two competing effects:
Calcium adsorption onto calcite surface rises the zeta potential and consequently the electrostatic repulsion, while increasing salt concentration induces stronger electrostatic screening.
We also demonstrate that the addition of sodium hydroxide completely screens the surface charge and leads to a more rigid paste.
A second important result is that carbonation of the calcite suspensions by the atmospheric CO$_2$ leads to a convergent high elasticity of the colloidal gels, whatever their initial value, also well rationalized by DLVO theory and resulting from a decrease in zeta potential and in surface charge density.

\end{abstract}

\end{frontmatter}

\section{Introduction}

Calcium carbonate, including its most stable polymorph, calcite, is one of the most abundant mineral on Earth. 
It is found in limestone, chalk and marble, and has been used for thousands of years as construction material.
As a major constituent of rock reservoirs, it is also
of crucial importance for oil recovery or CO$_{2}$ sequestration~\cite{Knauss2005}.
Finally, ground or precipitated calcite particles are used in various industries (cement, paper, etc.) as fillers or raw materials \cite{Benachour2008}. 

Related to these applications, rock-fracture \cite{royne_experimental_2011}, oil dewetting \cite{rezaei_gomari_effect_2006} or flow of colloidal suspensions \cite{pourchet2013chemistry} are macroscopic phenomena largely influenced by mineral surfaces and their interactions through aqueous solutions.
The solution physico-chemistry is influenced by the solubility product of the solid phases, including calcium carbonate, which in turn modify surface properties.
For instance, calcium and carbonate sites present at the calcite surface get hydrated {respectively by  the OH$^-$ and H$^+$} ions of the solution.
They also form complexes with, e.g. calcium and carbonate ions of the solution \cite{stipp1991structure,wolthers_surface_2008,heberling2011structure}.
These so-called potential determining ions thereby modify the calcite surface charge and zeta potential~\cite{al2017zeta}.

To characterize surface forces between calcite planes immersed in aqueous solutions, measurements based on atomic force microscopy \cite{pourchet2013chemistry, royne_repulsive_2015, diao_molecular_2016, javadi_adhesive_2018} and surface force apparatus \cite{dziadkowiec_surface_2018} have been undertaken in the recent years, but no consensus has emerged yet on the nature of dominant interactions between calcite surfaces.
These interactions are for example not easily described by the classical Derjaguin-Landau-Verwey-Overbeek (DLVO) model~\cite{israelachvili_intermolecular_1992}.
In particular, nano\-me\-tric-\-range repulsions have been evidenced \cite{royne_repulsive_2015, diao_molecular_2016} and attributed to the so-called repulsive secondary hydration forces originating from the compression and dehydration of hydrated counterions in the vicinity of the solid surfaces \cite{donaldson_developing_2015}.
However, experimentally, the reactivity-induced roughening of the calcite surface can also lead to an apparent repulsion \cite{dziadkowiec_surface_2018} and discriminating between the two effects is challenging.
Besides, additional non-DLVO attractive forces induced by  ion-ion correlation  are suspected to play a role in calcite surface interactions at high ionic strength \cite{javadi_adhesive_2018}.

A different approach to investigate mineral interactions is to study the mechanical responses of mineral suspensions as a function of the physico-chemical conditions.
The link between microscopic interactions and rheological properties of dense suspensions has indeed been investigated both theoretically \cite{hunter1968dependence,firth1976flowIII,shih1999elastic,flatt2006yodel} and experimentally \cite{scales1998shear,zhu2016yield,castellani2013ions,gossard2017rheological,leong1993rheological,friend1971plastic,firth1976flowII,gustafsson2000influence,kosmulski1999correlation}.
In particular, previous works focused on the relation between the yield stress $\sigma_y$ and the zeta potential $\zeta$~\cite{hunter1968dependence,firth1976flowIII,shih1999elastic,flatt2006yodel,leong1993rheological,friend1971plastic,firth1976flowII,gustafsson2000influence,kosmulski1999correlation} in the limit of small electrostatic repulsion.

In this study, we investigate the interactions between calcium carbonate surfaces in aqueous solutions  by performing macroscopic elasticity measurements of  suspensions of calcite colloids, combined with zeta potential measurements and chemical speciation calculations. 
Due to the specific shape and size of our calcite colloids, the pure calcite paste behaves as a typical colloidal fractal gel \cite{liberto2017elasticity}, revealing attractive interactions between particles.
By adding specific ions such as calcium, it is possible to increase efficiently the electrostatic repulsion, beyond the small repulsion regime.
By varying the solution physico-chemistry of this reactive suspension (i.e. ionic strength and ions content), we go from strongly attractive to almost repulsive systems. Nevertheless, all suspensions have an elastic-like behavior, spanning two orders of magnitude in shear modulus. 
We show a direct correlation between the paste elastic modulus and the DLVO energy barrier, which is not only tuned by the zeta potential but also by the Debye length, hence the ionic strength which is here varied from 1 to 100~mM.
We thereby obtain hints on the interaction forces at play at the microscopic scale between calcium carbonate particles in an aqueous environment.

More precisely, we have investigated how the addition of calcium hydroxide (Ca(OH)$_{2}$) or sodium hydroxide (NaOH) to the calcite suspension modifies its mechanical properties.
Calcium hydroxide has been chosen because calcium is a potential-determining ion of calcium carbonate 
\cite{pourchet2013chemistry,foxall1979charge,pierre1990calcium,huang1991adsorption, nystrom2001influence} inducing an increase of {the} {positive} zeta potential.
On the contrary, so\-dium hydroxide lowers the $\zeta$ potential.
Regarding rheological properties, we have focused on elastic modulus measurements and followed continuously the evolution of the paste with time.

Our first main result is that carbonation of the paste by the atmospheric CO$_2$ leads to a convergent high rigidity of the pastes, whatever their initial value.
Our second main result is that the initial elasticity of the calcite suspension exhibits a non-monotonous behavior with the concentration of calcium hydroxide, due to a crossover between two competing effects:
calcium adsorption rises the zeta potential while increasing salt concentration induces higher electrostatic screening.
Oppositely, the addition of sodium hydroxide strongly decreases the electrostatic repulsion and in\-crea\-ses the elastic modulus.

\section{Materials and Methods} \label{mem}

\subsection{Sample preparation}

We use Socal\,$31$ calcite powder (from Solvay, now available from Imerys)
with an average particle diameter of 70 nm, a density of 2710 kg/m$^3$ and a specific surface area of 17 m$^2$/g. The calcite particles are faceted as shown on the TEM image (Fig.\ref{TEM}).
To obtain a calcite suspension, the powder is homogeneously dispersed in various solutions using a vortex stirrer (Ultra Turrax TD300) at a mixing rate of 5800 rpm for 5 minutes.
The initial calcite volume concentration is fixed to $\phi=10$\,\%. 
To investigate the effects of simple ionic additives, we disperse  calcium hydroxide Ca(OH)$_{2}$ (concentration $c$ ranging from 3 to 50~mM) or  sodium hydroxide NaOH (concentration 94~mM) in deionized water. Both chemicals are from Sigma Aldrich. 

As we found that the addition of $3$~mM of Ca(OH)$_{2}$ to deionized water increases the pH of the initial solution up to 8, avoiding an initial calcite dissolution, $3$~mM is the smallest studied concentration of Ca(OH)$_{2}$. Moreover, we added this quantity of Ca(OH)$_{2}$  to the one containing NaOH.

For all the samples, pH values are measured with a pH-meter (Mettler-Toledo or Hana Edge), right after the beginning of the rheological test.
The reproducibility range of the pH values is $\pm0.2$.

\subsection{Rheological measurements}

In order to investigate the role of ionic additives on the shear elastic modulus of the calcite pastes, we use the following protocol.
The measurements are performed with a stress-controlled rotational rheometer (Anton Paar MCR 301) in a plate-plate geometry   at room temperature. The upper and lower plate diameters are 36 and 64~mm, respectively.
The gap  width is fixed at 1~mm and the plates are covered with sand paper (roughness 46~$\mu$m) in the aim to make slippage at the wall negligible~\cite{liberto2017elasticity}.
The measurements consist in two steps.
A first pre-shear step consists in a 1 minute imposed shear rate of $\dot{\gamma}=10$~s$^{-1}$, in order to start from comparable initial conditions for each sample.
We then apply a constant deformation of $\gamma=0.01$\,\%  at  frequency  $f=1$\,Hz during~10h, for which we
measure the temporal evolution of the storage modulus $G'(t)$ of the sample.
The imposed deformation is small enough to remain in the linear regime \cite{liberto2017elasticity}.
In  particular, we extract the initial value of the linear storage modulus $G'(0)$. 
During all measurements, we maintain the sample in a moisture chamber. It prevents the calcite paste from drying but does not insulate it totally from the atmosphere.

\subsection{DLVO calculation}

\begin{figure}
\centering
\includegraphics[height=5cm]{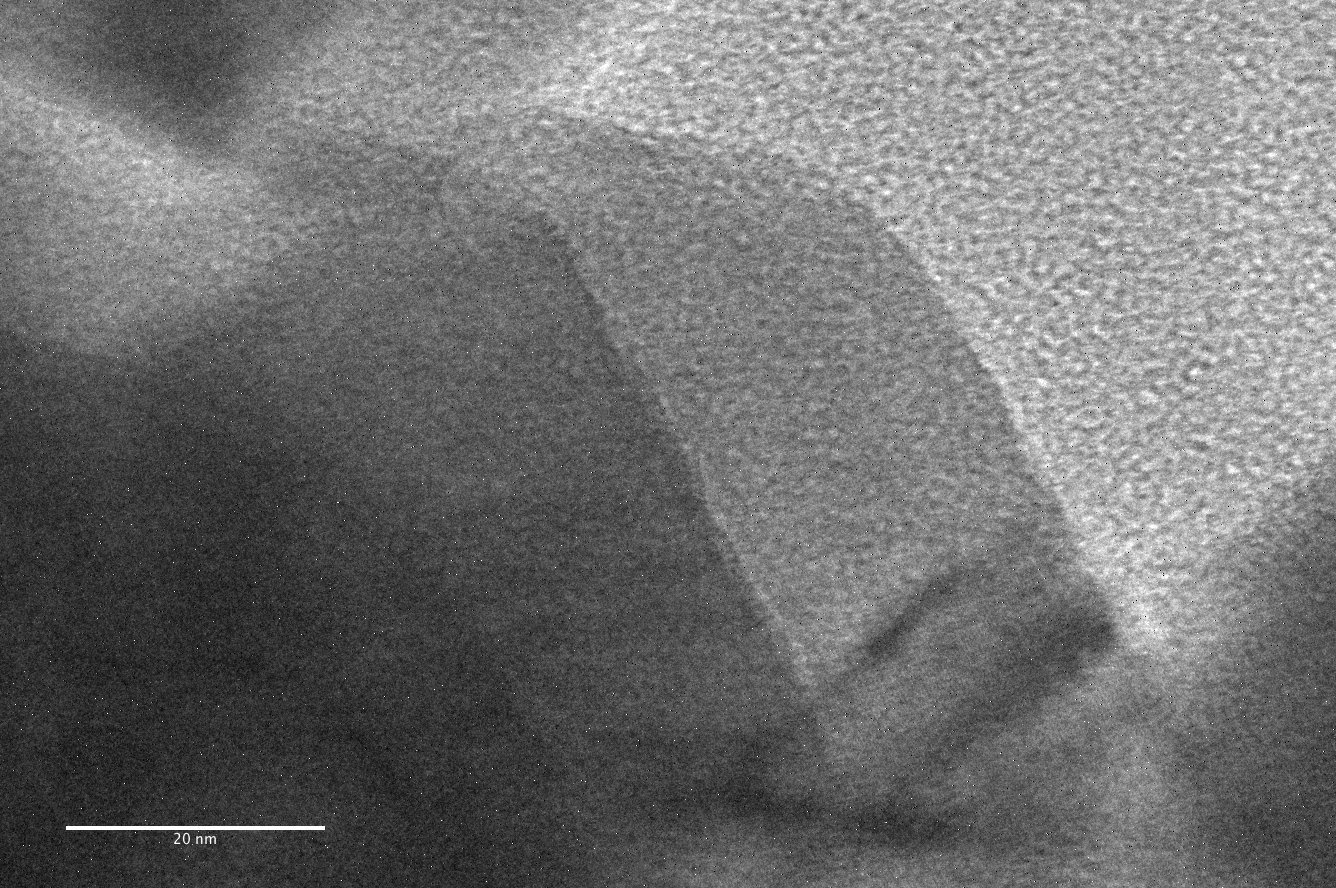}
\caption{TEM images of pure calcite powder showing plane surfaces (facets). 
The scale bar is equal to $20$\,nm. Image acquired at 120~kV.} 
\label{TEM}
\end{figure}

To characterize the strength of inter-particle interactions, we use the classical Derjaguin-Landau-Verwey-Overbeek theory 
\cite{israelachvili_intermolecular_1992}.
In this model, the interaction potential is the combination of two contributions: Van der Waals attraction and repulsion arising from the electrical double layer. 
As the studied particles have a nanometer-range size, they exhibit crystalline facets, as shown in  Fig. \ref{TEM}.
Therefore, the interaction between particles is considered here to proceed between parallel planes.

In this geometry and in the Debye-H\"uckel approximation, the DLVO potential per unit area reads:
\begin{equation}
\label{eq:DLVO}
W = -\frac{A}{12 \pi x^2} + \frac{2 \varepsilon}{\lambda_D} \zeta^2 \exp (-x/\lambda_D)
\end{equation}
with $x$ the distance between surfaces, $A$ the Hamaker constant, $\varepsilon$ the water permittivity, $\lambda_D$ the Debye length and $\zeta$ the zeta potential. 
The Van der Waals attraction is proportional to the Hamaker constant $A$, equal to $1.44 \cdot 10^{-20}\text{ J} \approx 3.6 \, k_BT$ for the calcite-water-calcite system \cite{bergstrom_hamaker_1997} and is inversely proportional to the distance squared.
The electrostatic repulsion has been calculated in the Debye-H\"uckel limit  corresponding to  small electrical potentials compared to $k_BT/e=25$ mV. In this limit, the repulsion varies as the square of the zeta potential $\zeta$, its range being given by the Debye length $\lambda_D$.

The competition between these two terms can lead to a non-monotonic potential profile as shown in Fig. \ref{DLVO_PP}. Attraction will be observed if the energy barrier is of the order or smaller than thermal agitation: $W_{\text{max}}\lesssim k_BT/a^2$.  

In order to calculate the repulsive term we need to know both $\zeta$ and $\lambda_D$.
The zeta potential can be measured directly on the paste as detailed in the following section. The Debye length is given by: 
\begin{equation}
\label{eq:lD}
\lambda_{D} = \sqrt{\frac{\epsilon k_{\text{B}}T}{2e^{2} {I}}}
\end{equation}
where  $I$ is the ionic strength, defined as 
%$I=\frac{1}{2}\displaystyle\sum_{i=1}^n c_{i} z_{i}^{2}$
$I=\sum_{i=1}^n c_{i} z_{i}^{2} / 2$ with $c_{i}$ and $z_{i}$ the concentration and valence of all species in solution.
The ionic strength is calculated from the chemical speciation of calcite suspensions, as explained in \Cref{minteq}.

\subsection{$\zeta$ potential measurements}

Zeta potential measurements are carried out directly on the concentrated calcite suspensions at $\phi$ = 10\% with an electroacoustic technique~\cite{dukhin1999electroacoustic,dukhin1999dynamic} by using a ZetaProbe instrument (DT-310 Dispersion Technology). The ultrasound wave generated by the ZetaProbe dipped in the suspension creates a pressure gradient that, in turn, generates an electrical current (Colloidal Vibration Current, CVI) detected by the probe itself. The electrical signal is converted into electrophoretic mobility and then in zeta potential according to the Smoluchowski model. Input parameters are the density (0.997~g/cm$^3$), viscosity (0.890~mPa s)  and relative permittivity (78.85) of the liquid medium (water) and the density of calcium carbonate.
%(2.75~g/cm$^3$) 
The particle size was set below 300~nm. The measurements have been repeated three times for each sample. The dispersion between these measurements is lower than 1\%. For the pure calcite, and the samples with concentrations of 30 and 50~mM of calcium hydroxide, two different samples were tested. The two measurements were reproducible with precision less than 1.5~mV.

\subsection{Chemical speciation}\label{minteq}

In order to quantify the ionic strength, the last unknown value in the DLVO calculation, the full ionic composition of the suspending fluid is calculated with the speciation freeware Visual MINTEQ 
\cite{minteq}.

\subsubsection{Calcite equilibrium reactions}\label{min_reactions}

The carbonate equilibria depend strongly on the pH and on the dissolved carbon dioxide. For pH above 10, dissolved CO$_2$ is mostly in the form of carbonate CO$_3^{2-}$ and for pH below 10 in the form of bicarbonate HCO$_3^{-}$.
Moreover, calcite once dissolved in water produces several chemical species such as: H$_{2}$CO$_{3}$ (carbonic acid), HCO$_{3}^{-}$, CO$_3^{2-}$, Ca$^{2+}$, CaHCO$_{3}^{+}$, CaOH$^{+}$, Ca(OH)$_{2\text{(aq)}}$ and CaCO$_{3\text{(aq)}}$ involved in the following reactions \cite{somasundaran1967zero}:
\\
\\
\schemestart
\label{a}
($a$)\hspace{0.2cm}
\chemfig{Ca{CO}_{3(s)}}
\arrow{<=>}
\chemfig{Ca{CO}_{3(aq)}}
\schemestop  
, $K{_a} = 10^{-5.09}$
\\
\schemestart
\label{b}
($b$)\hspace{0.2cm}
\chemfig{Ca{CO}_{3(aq)}}
\arrow{<=>}
\chemfig{Ca^{2+}} + \chemfig{CO_3^{2-}}
\schemestop  
, $K_b = 10^{-3.25}$
\\
\schemestart
\label{c}
($c$)\hspace{0.2cm}
\chemfig{CO_3^{2-}} + \chemfig{H_2O}
\arrow{<=>}
\chemfig{HCO_3^{-}} + \chemfig{OH^{-}}
\schemestop  
, $K_c = 10^{-3.67}$
\\
\schemestart
\label{d}
($d$)\hspace{0.2cm}
\chemfig{HCO_3^{-}} + \chemfig{H_2O}
\arrow{<=>}
\chemfig{H_2CO_3}  +  \chemfig{OH^{-}}
\schemestop  
, $K_d = 10^{-7.65}$
\\
\schemestart
\label{e}
($e$)\hspace{0.2cm}
\chemfig{H_2CO_3}
\arrow{<=>}
\chemfig{CO_{2(g)}} + \chemfig{H_2O}
\schemestop  
, $K_e = 10^{1.47}$
\\
\schemestart
\label{f}
($f$)\hspace{0.2cm}
\chemfig{Ca^{2+}} +  \chemfig{HCO_3^{-}} 
\arrow{<=>}
\chemfig{CaHCO_3^{+}} 
\schemestop  
, $K_f = 10^{0.82}$
\\
\schemestart
\label{g}
($g$)\hspace{0.2cm}
\chemfig{CaHCO_3^{+}} 
\arrow{<=>}
\chemfig{H^{+}} + \chemfig{Ca{CO}_{3(aq)}}
\schemestop  
, $K_g = 10^{-7.90}$
\\
\schemestart
\label{h}
($h$)\hspace{0.2cm}
\chemfig{Ca^{2+}} +  \chemfig{OH^{-}}
\arrow{<=>}
\chemfig{CaOH^{+}} 
\schemestop  
, $K_h = 10^{1.40}$
\\
\schemestart
\label{i}
($i$)\hspace{0.2cm}
\chemfig{CaOH^{+}} +  \chemfig{OH^{-}}
\arrow{<=>}
\chemfig{Ca{(OH)}_{2(aq)}} 
\schemestop  
, $K_i = 10^{1.37}$
\\
\schemestart
\label{j}
($j$)\hspace{0.2cm}
\chemfig{Ca{(OH)}_{2(aq)}} 
\arrow{<=>}
\chemfig{Ca{(OH)}_{2(s)}} 
\schemestop  
, $K_{j} = 10^{2.45}$
\\
\\
In particular, the calcite dissolution is described by reactions $(a)$ and $(b)$, the carbonate equilibria by reactions $(c)$ to $(e)$ and the calcium hydroxide dissolution by reactions $(h)$ to $(j)$.

Note that in our experiments we add calcium hydroxide  up to 50~mM.
The solubility limit of solid calcium hydroxide is around 20~mM~\cite{green2007perry}. 
Above this value, there  initially remains some solid calcium hydroxide Ca(OH)$_{2\text{(s)}}$, also called portlandite.

The above reactions can be recast to describe the carbonation of portlandite, i.e. its transformation into calcite
\\
\\
\schemestart
\label{k}
($k$)\hspace{0.1cm}
\chemfig{Ca{(OH)}_{2(s)}} + \chemfig{CO_{2(g)}}
\arrow{<=>}
\chemfig{CaCO_{3(s)}} + \chemfig{H_2O}
\schemestop  
, $K_k = 10^{13}$
\\
\\
This implies that the transformation of solid portlandite to solid calcite is thus favored as long as the activity of CO$_2$ in the air $a_{\text{CO}_2}$ is larger than $K_k^{-1} = 10^{-13}$, which is the case for atmospheric carbon dioxide ($a_{\text{CO}_2}$ = 0.00038).

Once portlandite is totally dissolved, further dissolution of CO$_2$ from the atmosphere leads to a pH decrease ---reactions $(c)$ to $(e)$--- while the carbonate ions react with calcium to precipitate as calcite ---reactions $(a)$ and $(b)$.

\subsubsection{Procedure for chemical speciation}

The chemical composition of our ionic solutions at equilibrium with calcite is calculated with the speciation software Visual MINTEQ~\cite{minteq} using the following procedure.

For all systems, calcite is imposed as an \textit{infinite solid phase}. 
Calcium hydroxide is incorporated by one of the two following equivalent procedures.
We can specify the corresponding concentrations of calcium and hydroxide ions, and add portlandite as \textit{possible solid phase} to allow Ca(OH)$_2$ to precipitate. Equivalently, we can incorporate portlandite directly as \textit{finite solid phase}. 

Then CO$_2$ is inserted as CO$_3^{2-}$ component (input) in order to obtain  the pH measured experimentally
ranging from 9 to 13, signature of the degree of advancement of the carbonation reaction $(k)$ in our sample.
MINTEQ then calculates the ionic concentrations to satisfy the above chemical equilibria $(a)$ to $(j)$, water dissociation as well as charge balance. Imposing the CO$_2$ pressure at an intermediate value between 0 and the atmospheric pressure is also possible and yields equivalent results. 

\section{Results}

\subsection{Time variation of shear elasticity}

\begin{figure}
\centering
\includegraphics[width=\columnwidth]{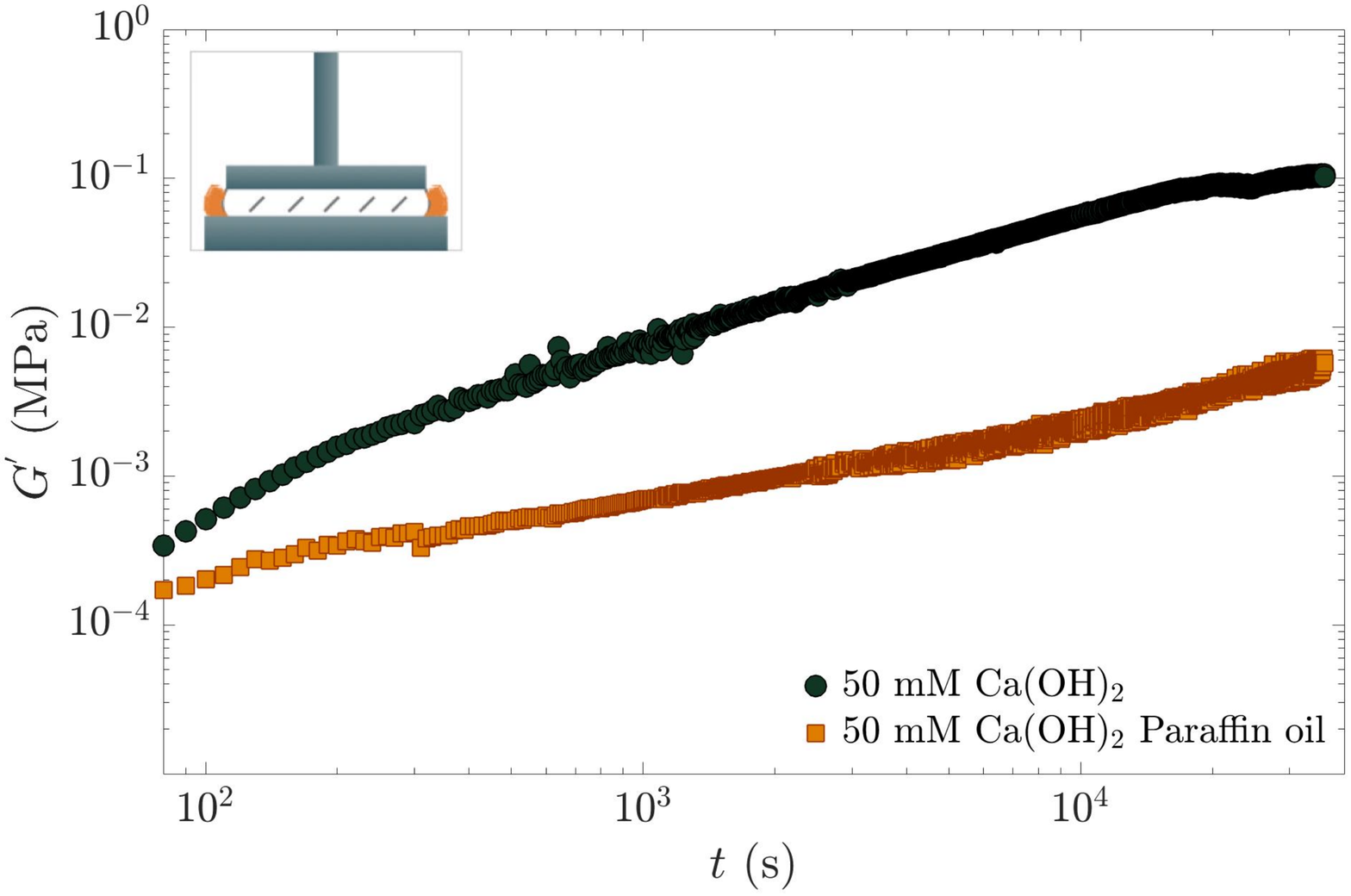}
\caption{Time evolution of the storage modulus of calcite suspension with $50$~mM of Ca(OH)${_2}$ for $\phi=10$\,\%.
In one case, paraffin oil is added around the plate-plate geometry to reduce the contact with the atmospheric CO$_{2}$ (see inset).}
\label{G_t}
\end{figure}

Fig. \ref{G_t} shows the temporal evolution of the elastic storage modulus $G'(t)$ for a sample containing $c =50$~mM of calcium hydroxide.
We observe that $G'$ increases with time by two orders of magnitude within a few hours.
Given the physico-chemistry of aqueous solutions containing calcite, we suspect that this temporal evolution is the consequence of the dissolution of atmospheric CO$_2$ (reaction  $(e)$), which modifies the solution physico-chemistry.
To test this hypothesis, we performed the same experiment but surrounded the sheared sample with paraffin oil, as sketched in the inset of Fig. \ref{G_t}, in order  to limit the contact with air.
We indeed observe that the sample with paraffin oil exhibits a slower time evolution of the storage modulus, with a final value at least one order of magnitude lower than the one of the sample in direct contact with the atmosphere, thus confirming the role of atmospheric CO$_2$ on the aging of the paste.

\begin{figure}%[ht]
\centering
\includegraphics[width=\columnwidth]{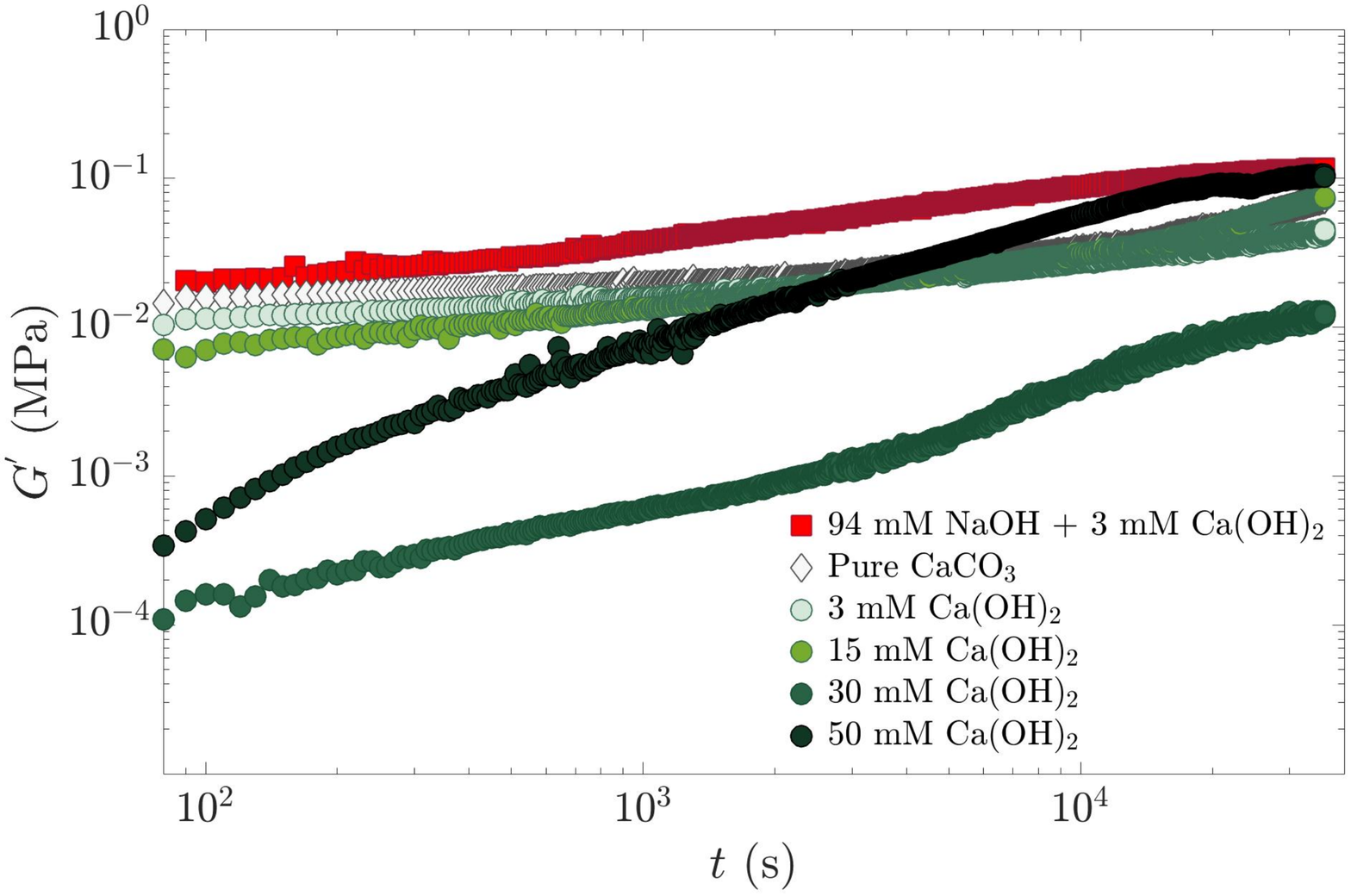}
\caption{Time evolution of the storage modulus $G'$ of calcite suspensions of concentration $\phi=10$\,\%, for increasing Ca(OH)${_2}$ concentrations and a sample containing sodium hydroxide.} 
\label{G_t_all}
\end{figure}

We now turn to the influence of the chemistry of the initial solution used to disperse calcite.
Fig. \ref{G_t_all} shows the temporal evolution of the storage modulus of  calcite pastes prepared with various concentrations of calcium hydroxide $c$ or with a high concentration of sodium hydroxide.
Two behaviors are observed: pure calcite paste and pastes containing sodium hydroxide or $c=3$ and 15~mM of calcium hydroxide present a slow increase of $G'$ with time, whereas suspensions with $c=30$ and 50~mM show a strong evolution. We also find that all the samples seem to converge on a timescale of $10^5$~s to a similar shear elastic modulus, of the order of 0.1 MPa.

In contrast, $G'(0)$, the initial value of $G'(t)$, shows a large variability with the nature and concentration of added ions.
More precisely, the initial elastic modulus decreases by two orders of magnitude as the calcium hydroxide concentration increases up to $c=30$~mM, before increasing as $c$ goes from 30 to 50~mM. 
We thus observe a non monotonic trend of $G'(0)$ with $c$.

To  interpret these observations, we now study in more detail the chemical composition of the dispersing solutions.

\subsection{Interparticle forces}

\subsubsection{Chemical speciation}

A complete chemical speciation is calculated for each sample with the software Visual MINTEQ using the protocol presented in \Cref{minteq}. 
For various initial calcium hydroxide concentrations $c$, results of the most relevant parameters are collected in \Cref{tab:minteq_Ca}. 

\begin{table*}
 \centering
   % \begin{tabular}{lrrrrrrrrr}
    \begin{tabular}{r|cc|cc|cccc|cc}%{0.9\textwidth}
    
    \  & \multicolumn{2}{c|}{No atmosphere} & \multicolumn{2}{c|}{atmospheric pressure} &  \multicolumn{4}{c|}{{initial} experimental condition} &  \multicolumn{2}{c}{calcite vol$\%$ change} \\
   \hline 
        {$c$ (mM)}  & \multicolumn{1}{c}{pH} & \multicolumn{1}{c|}{{[Ca$^{2+}$]} (mM)} & \multicolumn{1}{c}{pH} & \multicolumn{1}{c|}{{[Ca$^{2+}$]} (mM)} & \multicolumn{1}{c}{pH$_{\text{meas}}$} & \multicolumn{1}{c}{{[Ca$^{2+}$]} (mM)} & \multicolumn{1}{c}{$I$ (mM)} & \multicolumn{1}{c|}{$\lambda_D$ (nm)} &  \multicolumn{1}{c}{$\Delta\phi_0$ (\%)}& \multicolumn{1}{c}{$\Delta\phi_f$ (\%)} \\
         \hline  
    0  &  9.9  & 0.12 & 8.2  & 0.51 & 8.9  & 0.24 & 0.73 & 11.1 &-0.01 & 0.9 \\
    3  & 11.7  & 2.8  & 8.2  & 0.51 & 10.3 & 0.15 & 0.47 & 13.8 & 0.1 & 1.0 \\
    15 & 12.4  & 11.8 & 8.2  & 0.51 & 11.0 & 0.55 & 1.7  & 7.3  & 0.5 & 1.4 \\
    30 & 12.4  & 14.2 & 8.2  & 0.51 & 11.8 & 3.3  & 10.2 & 3.0  & 1.0 & 1.8 \\
    50 & 12.4  & 14.2 & 8.2  & 0.51 & 12.4 & 13.0 & 43.2 & 1.4  & 1.1 & 2.4\\
     \lasthline
  \end{tabular}
  \caption{MINTEQ chemical speciation for $\phi=10$\,\% samples {and} increasing {initial Ca(OH)$_2$} { concentration} $c$.
Values of pH and calcium ion concentration {[Ca$^{2+}$]} are reported for the two extreme theoretical cases (system without atmosphere and in equilibrium with the atmosphere) and for our experimental conditions. {$\Delta\phi_0$ (resp. $\Delta\phi_f$) is the variation in calcite volume fraction between the system at the initial experimental condition (resp. at equilibrium with atmosphere) and without atmosphere. $\Delta\phi<0$ correspond to dissolution and $\Delta\phi>0$ to precipitation.} {The symbol pH$_{\text{meas}}$ (respectively pH)} corresponds to the measured (respectively calculated) pH}.
  \label{tab:minteq_Ca}%
\end{table*}

We report the values of pH and calcium ion concentration {[Ca$^{2+}$]}  in the two extreme {theoretical} cases: a system without atmosphere and {a system} in equilibrium with the atmosphere (pCO$_2$ = 0.00038 atm). We also report the pH measured  at the beginning of the rheological measurement, denoted pH$_\text{meas}$, and the corresponding calculated {[Ca$^{2+}$]}. The measured pH is intermediate between the pH values reached in the two extreme cases, yet closer to the no atmosphere one, indicating that the exchange with CO$_2$ is limited during sample storage before the experiments.

We also note that the theoretical pH and [Ca$^{2+}$] of calcite paste with $c=30$ or $50$ mM are identical with and without atmosphere. This is because these concentrations are above the solubility limit of calcium hydroxide (20~mM).

The ionic strength $I$ given by Minteq from pH$_\text{meas}$  is also reported in Table~\ref{tab:minteq_Ca} as well as the corresponding Debye length $\lambda_D$ (Eq.~\ref{eq:lD}).

Finally, we show the {absolute variations in calcite volume fraction} $\Delta \phi_0$ at the beginning of the experiment and $\Delta \phi_f$  at long times.  $\Delta \phi_0$ corresponds to the difference of the solid volume fraction $\phi$ between the initial experimental conditions and chemical equilibrium without atmosphere. $\Delta \phi_f$ is calculated as the difference between chemical equilibria with and without atmosphere.

$\Delta\phi<0$ {(respectively $\Delta\phi>0$) corresponds to a dissolution (resp. precipitation)}  of { calcite}.
We find that calcite precipitation occurs in all samples, except for the pure calcite one, where limited dissolution initially takes place.

For the sample containing sodium hydroxide NaOH, the experimental chemical composition deduced from  pH$_\text{meas}=12.7$ is: [Ca$^{2+}$] = 1.8 $\cdot$ 10$^{-3}$~mM, $I = 90$~mM and $\lambda_D = 1$~nm.

\subsubsection{Zeta potential measurements}

\begin{figure}%[ht]
\centering
\includegraphics[height=5.3cm]{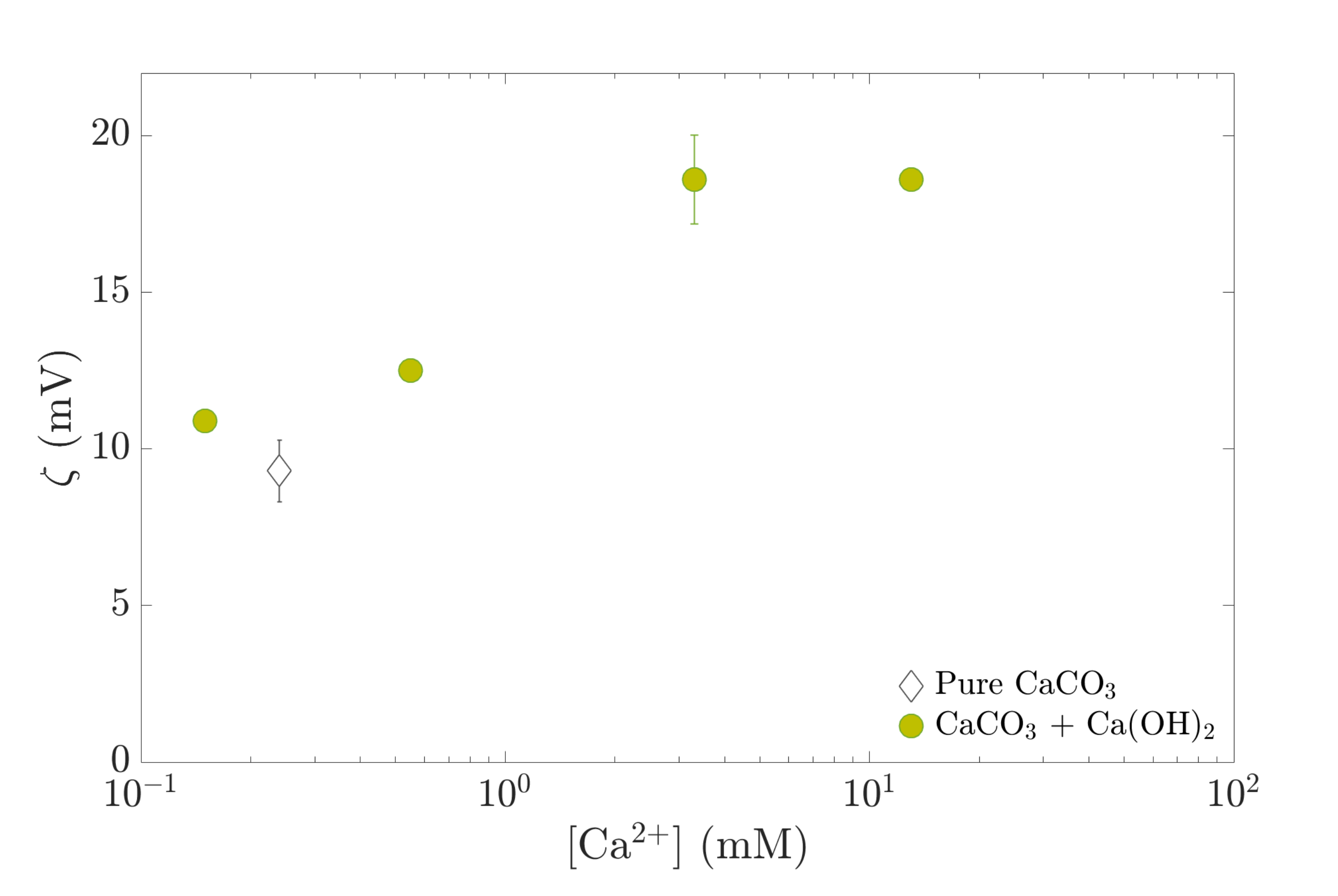} 
\caption{Zeta potential $\zeta$ as a function of the concentration of calcium ions Ca$^{2+}$ deduced from the speciation calculation (\Cref{tab:minteq_Ca}).
The calcium content is modified by changing the Ca(OH)$_2$ concentration $c$.}
\label{z_caoh2}	
\end{figure}

In Fig. \ref{z_caoh2}, zeta potential values are plotted as a function of the corresponding Ca$^{2+}$ concentration in the suspending solution. The calcium concentration is deduced from the speciation calculation and increases with the initial calcium hydroxide concentration $c$.

{We observe that} the zeta potential, {initially positive,} increases, then saturates with the {concentration} of calcium ions. {This confirms that calcium is a potential determining ion for calcite}~\cite{foxall1979charge,pierre1990calcium}, {as it adsorbs preferentially on the calcite surface} \cite{pourchet2013chemistry}.
\\

\subsubsection{DLVO calculations}
\label{DLVOcalc}

\begin{figure}%[ht]
\centering
\includegraphics[height=6cm]{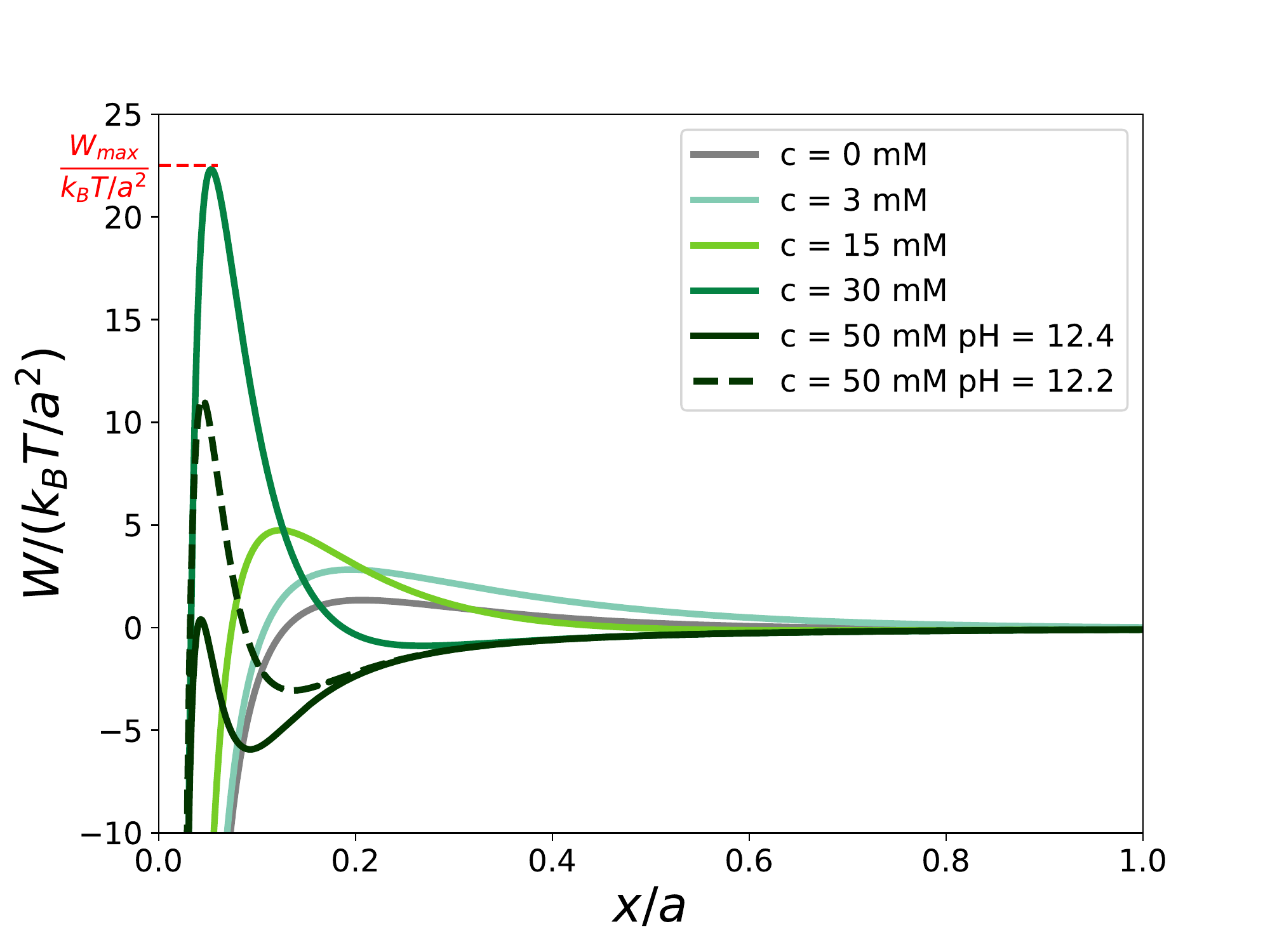} 
\caption{Plane-plane normalized {DLVO} interaction potentials as a function of the distance normalized by the particle size $a$.
The various curves correspond to calcite suspensions containing an increasing initial concentration of calcium hydroxide $c$.
For $c = 50$~mM, two curves at different pH are plotted to show the sensitivity of the interaction to the pH value (measured $\pm$0.2).
In red, we highlight the maximum value of the interaction potential $W_{\text{max}}$ for $c = 30$~mM.}
\label{DLVO_PP}
\end{figure}

Once known the ionic strength, from the chemical speciation, and the zeta potential, we calculate the DLVO interaction potential between two infinite calcite planes.
Fig. \ref{DLVO_PP} shows the interaction potential per unit area as a function of the interparticle distance, calculated for various values of the initial calcium hydroxide concentration $c$.

For pure calcite the potential is weakly attractive with a barrier of the order of $k_B T/a^2$.
Increasing $c$, the repulsion barrier increases to around {$22~k_BT/a^2$} for $c=30$~mM, then drops down to $0.4~k_BT/a^2$ for $c=50$~mM.
{Note that} the potential and consequently the energy barrier estimated at 50~mM  are very sensitive to the pH value used for chemical speciation, due to large changes in the ionic strength.
To illustrate this, we plot in Fig. \ref{DLVO_PP} two curves for  $c=50$~mM: a solid line for pH$=12.4$ (measured value, corresponding to $W_{\text{max}} \approx 0.4~k_BT/a^2$) and a dashed line for pH$ = 12.2$ ($W_{\text{max}}\approx 11~k_BT/a^2$). We find that a change in pH comparable to the experimental uncertainty turns the interaction from a repulsive to an attractive one for this concentration.
{This point will be discussed in detail in the following section.}

\begin{figure}%[ht]
\centering
\includegraphics[height=6cm]{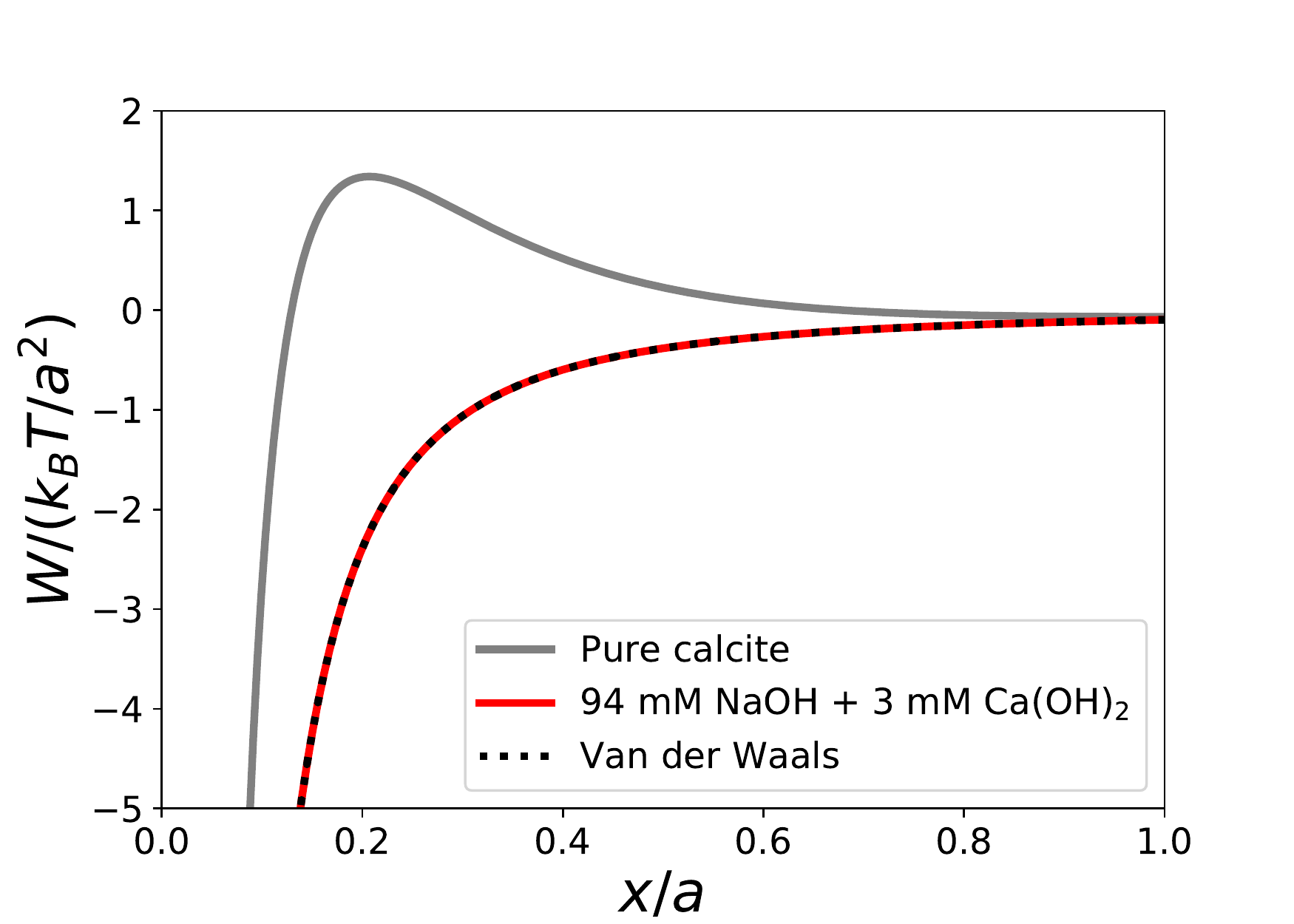} 
\caption{{Plane-plane DLVO} interaction potential for pure CaCO$_{3}$ (gray) and CaCO$_{3}$ with {NaOH} (red).
The dashed line shows the contribution of van der Waals attraction.}
\label{DLVO_PP_NaOH}	
\end{figure}

Fig. \ref{DLVO_PP_NaOH} shows {the results of the DLVO calculation of} the interaction energy for pure calcite and calcite with {sodium hydroxide}.
In the latter case, attraction is  dominant and the energy profile  coincides with pure Van der Waals attraction.
{The vanishing of the electrostatic repulsion results from the low zeta potential ($\zeta=-2.4$~mV), inducing  a weak repulsion between the planes, and from the high ionic strength ($I=90$ mM), which screens the surface charges and consequently reduces the range of the repulsion ($\lambda_D \approx 1$~nm).}

\section{Discussion}

\subsection{Zeta potential}

Calcium Ca$^{2+}$ and carbonate CO$_3^{2-}$ are known as potential determining ions of calcite \cite{al2017zeta}.
Calcite surface sites are predominantly in their hydrated forms $\text{>CaOH}^{-(1-x)}$ and $\text{>CO}_3\text{H}^{+(1-x)}$  with $x \approx 0.25$~\cite{heberling2011structure, song_surface_2017}.
Potential determining ions form complexes  with these surface sites, and thereby modify the overall surface charge and potential.
At high calcium concentrations, its surface complexation is expected to be favored, leading to a zeta potential increase. On the contrary, increasing carbonate concentrations should lead to a decrease in zeta potential.

We observe in Fig.\ref{z_caoh2} that $\zeta$ indeed increases with $[\text{Ca}^{2+}]$, as more positively charged calcium ions adsorb on the surface. For $[\text{Ca}^{2+}] > 3$~mM, we observe that $\zeta$ saturates. This behavior and the order of magnitude of the measured potentials are consistent with previous measurements on calcite particles~\cite{pourchet2013chemistry}.

We also observe that the zeta potential of pure calcite is slightly lower than the one measured with calcium hydroxide. This can be attributed to  the role of the  potential determining anions CO$_3^{2-}$ and HCO$_3^{-}$. The corresponding concentrations are presented in \Cref{tab:carbonate}. 
For pure calcite, the main anion is HCO$_3^-$, which complexes onto calcite (although more weakly than carbonate CO$_3^{2-}$ \cite{heberling2011structure,song_surface_2017}) and thereby makes $\zeta$ decrease.
{In all the solutions containing calcium hydroxide}, the main anion is {no more bicarbonate but} hydroxide, which is not a potential determining ion, {and therefore does not affect the $\zeta$  potential.}

\begin{table}%[ht]
 \small
 \centering
   % \begin{tabular}{lrrrrrrrrr}
    \begin{tabular}{r|ccc}
    {$c$ (mM)}  & \multicolumn{1}{c}{[Ca$^{2+}]$ (mM)} & \multicolumn{1}{c}{[CO$_3^{2-}]$ (mM)} & \multicolumn{1}{c}{[HCO$_3^{-}]$ (mM)}\\ %\multicolumn{1}{c}{[OH$^{-}]$ (mM)} {PDI ratio}
         \hline  
    0  & 0.24  & 1.8 $\cdot$ 10$^{-2}$  & 4.2 $\cdot$ 10$^{-1}$ \\% 8.6 $\cdot$ 10$^{-3}$ \\ % 13.1
    3  & 0.15  & 2.8 $\cdot$ 10$^{-2}$  & 2.7 $\cdot$ 10$^{-2}$\\% 2.1 $\cdot$ 10$^{-1}$  \\ %& 5.2
    15 & 0.55  & 8.7 $\cdot$ 10$^{-3}$  & 1.6 $\cdot$ 10$^{-3}$ \\% 1.1   \\ %63.2
    30  & 3.3  & 2.3 $\cdot$ 10$^{-3}$  & 5.8 $\cdot$ 10$^{-5}$ \\% 6.9  \\ %1.4$\cdot$ 10$^{3}$
    50 & 13.0  & 1.1 $\cdot$ 10$^{-3}$  & 5.5 $\cdot$ 10$^{-6}$ \\% 30.1  \\ %1.2$\cdot$ 10$^{4}$
     \lasthline
  \end{tabular}
  \caption{Concentrations of  potential determining ions obtained by MINTEQ chemical speciation at the experimental conditions, {with} increasing Ca(OH)${_2}$ concentration $c$.
}%The potential determining ions (PDI) ratio is defined in the text.
  \label{tab:carbonate}%
\end{table}

In the sample with sodium hydroxide, the positive charge of Na$^{+}$ ions (86~mM) is balanced with the negative ions: OH$^{-}$ (52~mM), CO$_3^{2-}$ (13~mM) and Na{CO}$_{3}^{-}$ (8~mM). The negative  potential determining ion CO$_3^{2-}$ is much more concentrated than calcium (1.8 $\cdot$ 10$^{-3}$~mM). In the same time, sodium ions Na$^+$ also (weakly) bind to negatively charged surface sites \cite{heberling2011structure,song_surface_2017}. This results in a slightly negative value of $\zeta=-2.4$~mV.

\subsection{Validity of Poisson-Boltzmann approach}

The elastic behavior of our calcite paste, observed at small volume fraction, down to $5\%$ \cite{liberto2017elasticity}, suggests that attraction between calcite particles is significant and leads  to the formation of a gel.
The question we have to answer is the origin of the attraction. For chemically identical colloids, Van der Waals interaction induces attraction between colloids independently of the ion concentration, ion valence or particle surface charge \cite{israelachvili_intermolecular_1992}.
For highly charged colloids or polyelectrolytes in presence of multivalent ions, correlations of ion fluctuations along the surfaces can lead to attraction \cite{bloomfield1991condensation, rouzina1996macroion, butler2003ion, allahyarov2004attraction, delville1997monte, labbez2010colloidal}.
As it derives from fluctuations, this attraction cannot be described by mean-field Poisson-Boltzmann theory. Moreover it is enhanced by ions valence and surface charge \cite{butler2003ion, allahyarov2004attraction}. The importance of these two quantities on the validity of the Poisson-Boltzmann theory has been quantitatively evaluated by Netz \textit{et al.} using the non-dimensional parameter $\Sigma$ defined by \cite{attard1988beyond, netz2000beyond}:
$$ \Sigma =2\pi z^3\sigma_s \ell_B^2
$$ 
where $z$ is the ion valence, $l_B$ the Bjerrum length defined as $\ell_B=e^2/(4\pi\epsilon k_B T)$ and $\sigma_s$ the charge number per unit area (called charge density in the following). For $\Sigma <1$, ion-ion correlations are negligible and Poisson-Boltzmann  theory is valid \cite{netz2000beyond}.
However, for $\Sigma =2$ or higher values, ion-ion correlations {start} to be important and it is necessary to go beyond the mean-field theory.
In our calcite paste, the highest valence is $z=2$ and the Bjerrum length is $\ell_B=0.7$ nm. The most delicate parameter to estimate is the charge density $\sigma_s$ of the calcite particle. We estimate it from the zeta potential as:
$$\sigma_s\simeq \epsilon \zeta/(e\lambda_D ) $$
This relation between the potential and the charge density is obtained using Poisson-Boltzmann theory. For pure calcite paste, $\zeta =10$ mV, $\lambda_D=10$ nm, so that  $\sigma_s=5\times 10^{-3}$ charges/nm$^2$ corresponding to one charge every 10 nm$^2$.
Note that this charge density is very small compared to systems such as ADN or virus polyelectrolites \cite{bloomfield1991condensation, rouzina1996macroion} (typically 1 charge every nm$^2$) where ion-ion correlations are the dominant phenomena at small distances. 
Using this value, 
%those values,
we find that the non-dimensional parameter is $\Sigma =0.12$, i.e., one order of magnitude smaller than the limit $\Sigma =1$, underlying the relevance of the Poisson-Boltzmann theory that we used in our study.
However, estimating the surface charge density from the zeta potential is questionable due to ion condensation onto the surface \cite{manning1969limiting}. This approach can indeed underestimate the real surface charge  \cite{stipp1999toward, labbez2006surface,labbez2010colloidal}.  Combining experimental results and Monte-Carlo simulations, Labbez \textit{et al.} \cite{labbez2006surface} studied in detail the impact of ion condensation on the real charge density of calcium silicate hydrates and they showed that the surface charge can be underestimated at most by a factor~6.  If we take this factor into account, we find a non-dimensional parameter $\Sigma =0.7$  still smaller than 1. We therefore conclude that ion-ion correlations are not the relevant phenomena in our calcite paste and  that van der Waals interactions are at the origin of the attraction between calcite particle. This contrasts with recent AFM measurements suggesting a significant role of ion-ion correlations on interaction forces between calcite surfaces \cite{javadi_adhesive_2018}.

\subsection{Elastic modulus versus DLVO potential}
\subsubsection{Small repulsion regime}

In Fig. \ref{G_t_all}, we observe that the initial storage modulus of the fresh paste $G'(0)$  shows a minimum for the sample containing $c=30$~mM of Ca(OH)$_{2}$.

To interpret these data, we first turn to the celebrated models of 
Shih et al.~\cite{shih1999elastic} and Flatt \& Bowen~\cite{flatt2006yodel}, both relating the interparticle force and suspension structure to the mechanical properties of the paste.
These two models predict that the elastic modulus decreases linearly with $\zeta^{2}/\lambda_\text{D}$.
In our experiments, both $\zeta$ and $\lambda_\text{D}$ change with the concentration $c$ of calcium hydroxide, and the combined changes of both factors could lead to the non-monotonicity of $G'(0)$ with $c$ seen in Fig. \ref{G_t_all}.

\begin{figure}%[ht]
\centering
\includegraphics[height=6cm]{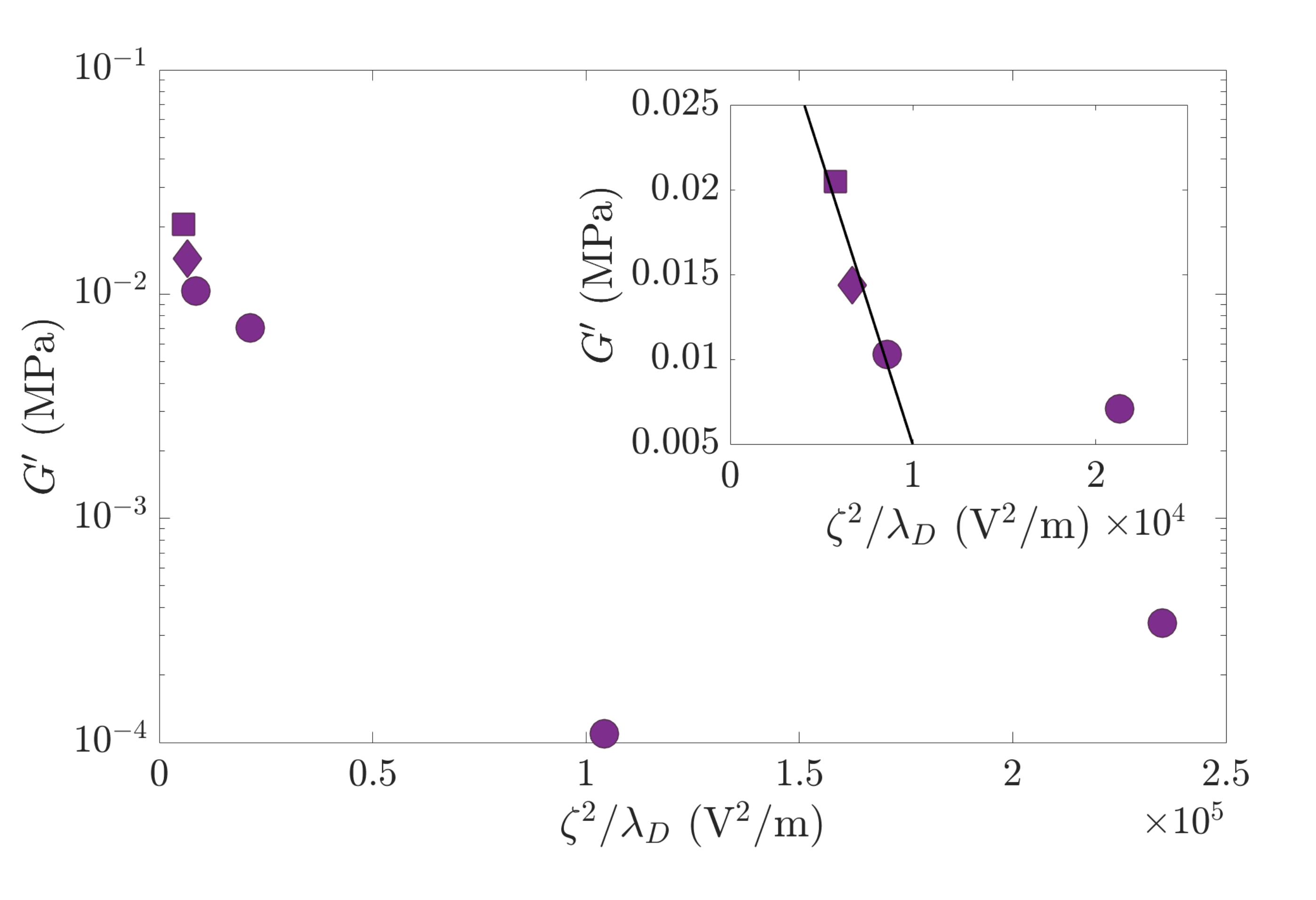} 
\caption{Initial storage modulus $G'(0)$ as function of $\zeta^{2}/\lambda_D$. All suspensions are collected, in particular calcite with NaOH (square), pure calcite (diamond) and calcite with Ca(OH)$_2$ (dots). In the inset is reported the linear fit of the first three points with the interaction model predictions by Shih et al. (1999)~\cite{shih1999elastic}.}
\label{G_fit}	
\end{figure}

Fig. \ref{G_fit} shows the evolution of $G'(0)$ with $\zeta^{2}/\lambda_\text{D}$.
We immediately see that these models are unable to explain the behavior of our paste.
As shown in the inset of Fig. \ref{G_fit}, only the first three points
---calcite with NaOH, pure calcite, calcite with 3 mM Ca(OH)$_2$---
decrease linearly with $\zeta^{2}/\lambda_\text{D}$.
For higher values of $\zeta^{2}/\lambda_\text{D}$, $G'(0)$ shows a non-linear variation.

In fact, this linear relation is valid only for systems showing small interparticle repulsion and logically fails for stron\-gly repulsive systems ---$c=15, 30$ and 50~mM--- appealing for a more complete modelization of the relation between rheological parameters and interparticle forces \cite{peng2012based}.

\subsubsection{Energy barrier}

\begin{figure}%[ht]
\centering
\includegraphics[height=5.5cm]{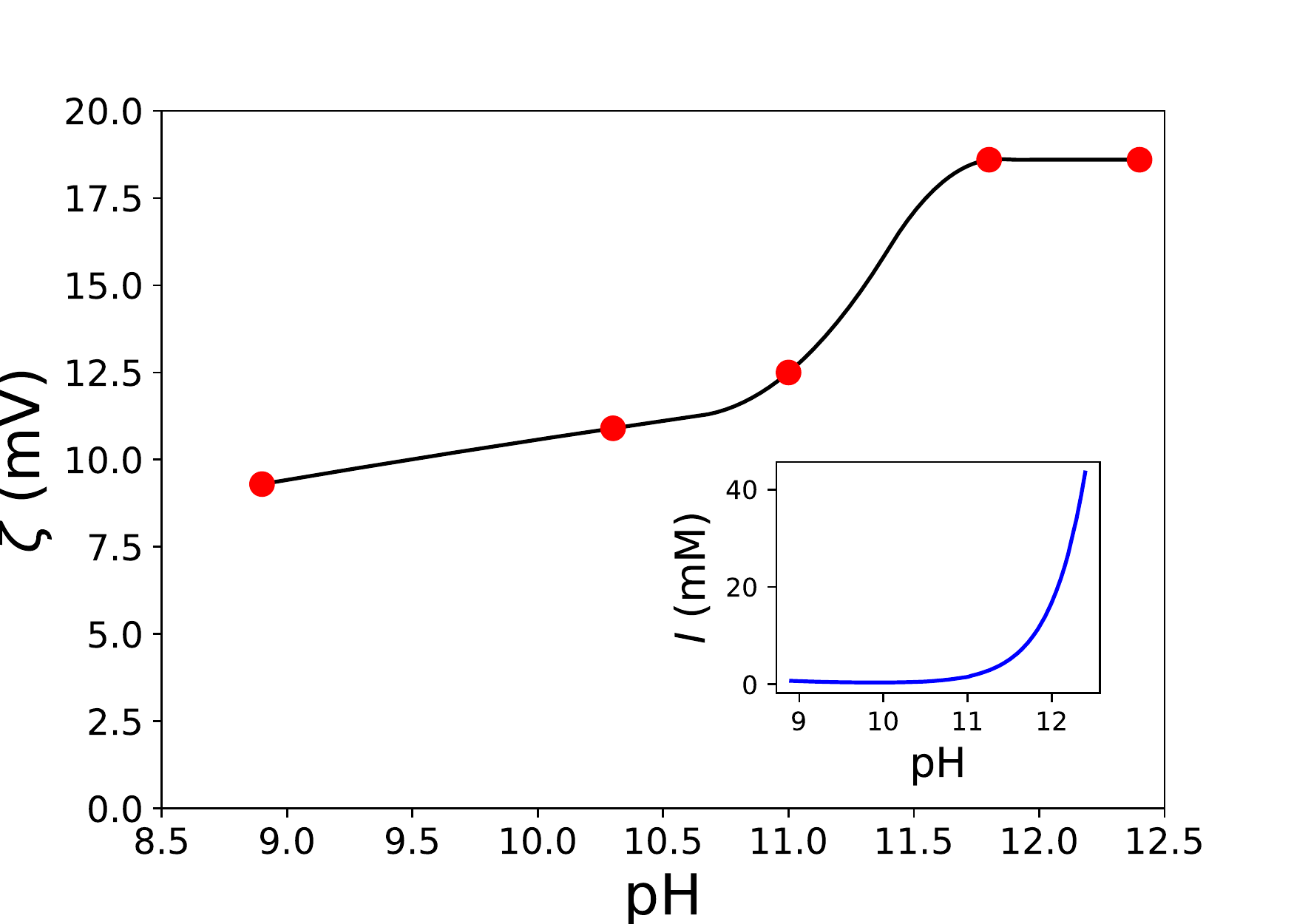} 
\caption{$\zeta$-potential as a function of pH. The line corresponds to an interpolation from experimental values (dots). Inset: ionic strength $I$ as a function of pH, calculated from chemical speciation.}
\label{Vmax_fit}	
\end{figure}

In order to go beyond the linear regime of small electrostatic repulsion, we use the DLVO computations developed in \Cref{DLVOcalc}.
These calculations make use of the zeta potential values of our suspensions, represented as a function of the pH in Fig. \ref{Vmax_fit}, and of its ionic strength $I$ (inset), from which the Debye length is deduced.
$I$ exhibits an extreme sensitivity on the pH in the very alkaline domain, whereas $\zeta$ shows only a mild increase with the pH.

\begin{figure}%[ht]
\centering
\includegraphics[width=8cm]{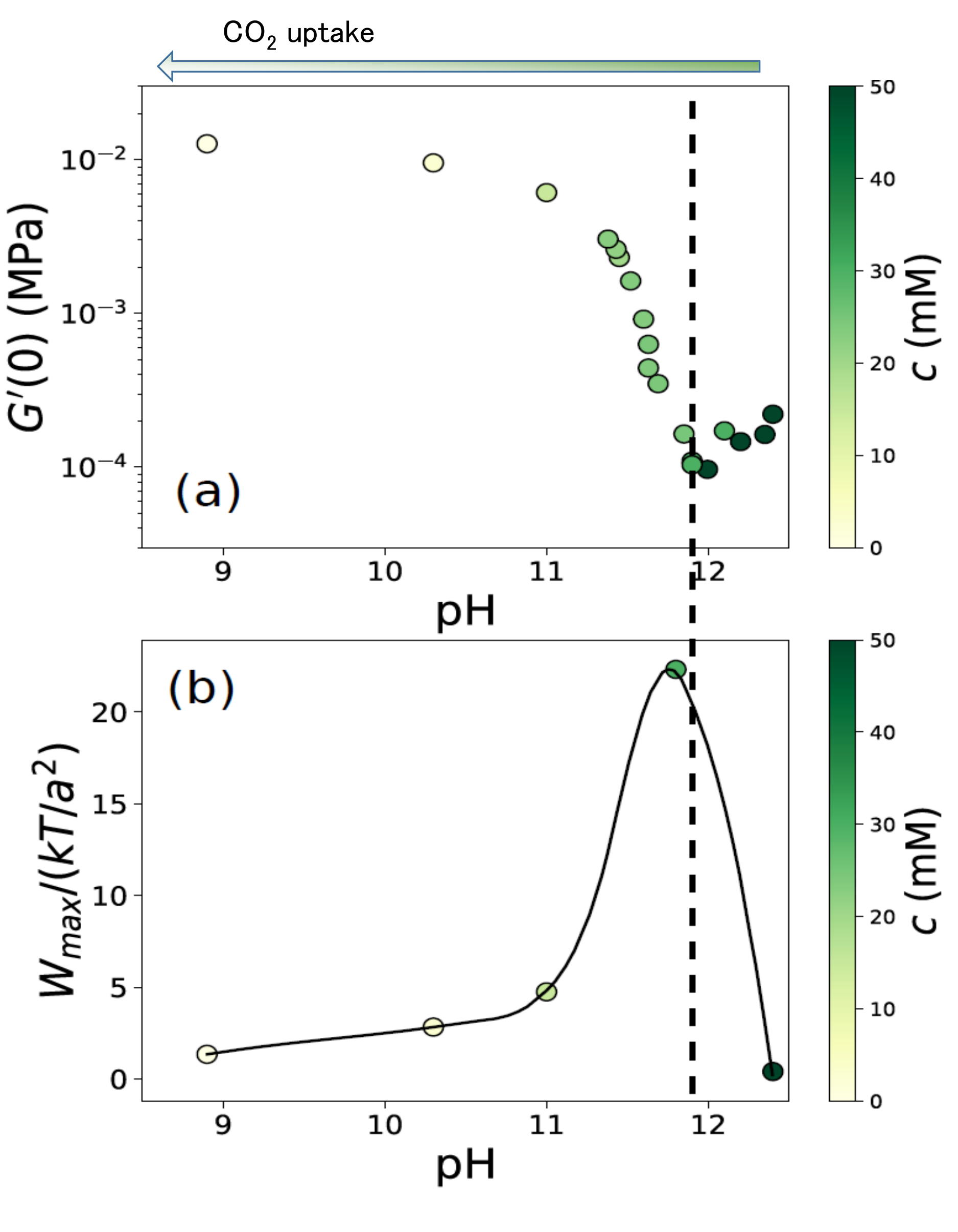} 
\caption{(a) Initial storage modulus $G'(0)$ as a function of pH. The different data points correspond to various $c$ shown by the colorscale. (b) $W_\text{max}/(k_BT /a^2)$ as a function of pH. The dots correspond to actual measurements of $\zeta$ while the continuous line is interpolated (see Fig.\ref{Vmax_fit}).The dashed line is a guide for the eye.}
\label{G_pH}	
\end{figure}

The effects of the increase of $\zeta$ and $I$ on the DLVO potential are opposite.
The former enhances the electrostatic repulsion whereas the latter reduces the double layer repulsion.
The consequent DLVO potential results from a trade-off between both.
To estimate the change of interaction nature with the calcium hydroxide concentration, we have computed the value of the DLVO potential maximum $W_\text{max}$  against pH, as shown in Fig. \ref{G_pH}b.
Its evolution is strongly non linear, with a sharp maximum at pH $\simeq$ 12.
The initial elastic modulus of the paste in the same pH range is presented in Fig. \ref{G_pH}a.
These data correspond to additional measurements at various $c$ in the range 0-50~mM.

The evolution of interaction potential and mechanical pro\-per\-ty are seen to be strongly correlated: The weaker the attraction, the softer the gel.
The $G'(0)$ minimum and $W_\text{max}$ maximum match perfectly at pH $\approx 12$.
Below this value, the $\zeta$ potential dominates, and makes the repulsion increase with the pH.
Above this threshold, the surface charge screening by the Debye layer preponderates and induces a decrease of the repulsion with the pH.
This strong agreement between the evolution of the microscopic interactions and their macroscopic counterpart demonstrates that simple ions added to a calcite paste with a millimolar concentration affect substantially its inter-particle interaction, which modifies its initial rigidity by orders of magnitude.

\subsubsection{Role of particle geometry}

\paragraph*{Particle anisotropy}

Our calcite particles have rhombohedral shapes. As a first approximation, we have therefore restricted our DLVO calculations to interactions between plane surfaces. However, the edges or vertices of the particles can also contribute to interparticle forces. 
Using Derjaguin approximation, the edge-facet or vertex-facet interaction potentials can be calculated as cylinder-plane or sphere-plane potentials, with a radius of curvature of $\sim 10$~nm estimated from TEM images (Fig. \ref{TEM}). We find that the amplitude of the interaction is largely reduced when going from plane-plane to edge-plane,  to vertex-plane. For instance, for the most repulsive system ($c = 30$~mM), the repulsion maximum goes from $22 k_BT/a^2$ to $5 k_BT/a$, to $0.2 k_BT$. While facets strongly repulse each other, particles can then attach via their vertices resulting in a percolated network, i.e. a gel, characterised by a non-zero elastic modulus.
This sensitivity of the interaction between faceted particles to their relative orientations could explain the very small value of critical strain that we measured on calcite paste \cite{liberto2017elasticity}.

\paragraph*{Particle roughness}

Upon growth or dissolution, the roughness of mineral surfaces is expected to increase \cite{deAssis2018}.
This results in an apparent repulsion between calcite surfaces measured in surface force apparatus experiments \cite{dziadkowiec_surface_2018}.
In our experiments with calcium hydroxide, it is likely that calcium carbonate precipitation, deriving from CO$_2$ uptake, proceeds through the growth of the primary particles.
However, after 10~hours, elasticity measurements of the precipitated ($c = 50$~mM) or slightly dissolved ($c = 0$~mM) calcite pastes are very close, suggesting limited roughening.

We believe this is due to the small size of our colloids. Mineral growth usually proceeds through step propagation but typical distances between steps are typically 1~$\mu$m, one order of magnitude larger than the size of our particles (70~nm) \cite{Zareeipolgardani2019}.
This would result in a limited growth-induced roughening of the calcite surfaces.

\subsection{Paste aging}

\subsubsection{Long-term equilibration}

Whatever the initial Ca(OH)$_2$ concentration, the elastic modulus of all the pastes converges toward the value of the pure calcite paste in the long run.
Therefore the uptake of atmospheric CO$_2$ by the suspension cancels progressively the effect of the dissolved portlandite.

This neutralization can be understood via the equilibrium equations introduced in \Cref{minteq}.
The CO$_2$ absorption induces pH decrease and calcite precipitation ---reactions $(a)$ to $(e)$---, which in turn leads to a [Ca$^{2+}$] reduction.

The pH decrease ---as soon as its value goes below 12--- causes a stiffening of the paste, as illustrated in Fig.\ref{G_pH}a.
As detailed in the previous section, this gain in rigidity finds its origin in the fall of the interparticle repulsion, originating in the $\zeta$ potential reduction.

In the same time, the calcite solid volume fraction increases due to precipitation. However, the observed evolution of $G'$ is not primarily due to this increase: For pure calcite, $G'$ increases by a factor 5, while for $c=30$~mM, $G'$ increases by two orders of magnitude. For both samples, the variation in solid volume fraction are comparable ($\Delta \phi_f - \Delta \phi_0 \approx 1\%$).

The steady state values of $G'$ for all initial portlandite concentration are expected to be very close, comparable to the elastic modulus of a pure calcite paste at pH 8.2, in equilibrium with the atmosphere. A small dispersion is expected due to differences in final solid volume fraction \cite{liberto2017elasticity}. \\

\subsubsection{Kinetics for carbonation}

The observed timescale for the paste stiffening is of the order of $10^4$~s (Fig.~\ref{G_t_all}). We here discuss what sets this timescale.

The kinetics for calcium hydroxide carbonation is ruled by the interplay between carbon dioxide diffusion in the air, diffusion of ionic species in solution (especially calcium, carbonate and hydroxyde), precipitation of calcite and, for $c\geq 20$~mM, dissolution of portlandite. We did not consider the nucleation of other crystalline forms of calcium carbonate (e.g. aragonite or vaterite) due to the high concentration of calcite with a large specific surface area (17 m$^2$/g). All other reactions are assumed to take place on much shorter timescales, less than 1~s \cite{mitchell_model_2010}. 

The geometry corresponding to the paste sample in the rheometer is a flat cylinder of height equal to the gap width 1~mm and of radius $R =18$~mm in contact with the atmosphere on the edges.

The diffusion coefficient of carbon dioxide in air is of the order of $10^{-5}$~m$^2$/s. Considering the gap width as the typical lengthscale of the problem, we find that a steady diffusive profile sets after 0.1~s in air, much faster than the observed change in elastic modulus.

The precipitation rate of calcite is of the order of $k_0 =10^{-6}$~mol/m$^2$/s (for a relative supersaturation index of 1) \cite{lakshtanov_inhibition_2011}. Given our calcite concentration, the calcite surface area per unit volume is $S= 5 \cdot 10^3$ m$^2$/L. Taking calcium hydroxide saturation $c_\text{max} \approx 20$~mM as a typical concentration, we find that the timescale for calcite precipitation is $c_\text{max}/(k_0 S) \sim 4$~s.

The dissolution rate of portlandite (solid calcium hydroxide) is of the order of $k_1 = 30\cdot 10^{-6}$~mol/m$^2$/s \cite{gay_local_2016}. We do not know the specific area of our calcium hydroxide, so we assume it is comparable to that of calcite $s =20$~m$^2$/g. We thus estimate the timescale for portlandite dissolution as $1/(k_1 M s) \approx 20$~s with $M=74$~g/mol the molar mass. 
Finally, the time $R^2/D$ for the ions to diffuse in the whole cylinder, of radius $R=18$ mm, is $\approx 3\cdot 10^{5}$~s with $D \sim 10^{-9}$ m$^2$/s, 
hence larger than the timescale for the $G'(t)$ increase. However, given the cylindrical sample geometry, the rheological response is dominated by that of the sample close to its edges. Taking a smaller typical length of 1~mm, we find a diffusive time $10^3$~s. The measured increase of $G'(t)$ indeed takes place on an intermediate timescale between the two estimates.

Finally, based on these orders of magnitude, we find that diffusion of ions in the solution is clearly the limiting rate for the carbonation dynamics, hence the temporal evolution of the shear modulus $G'(t)$.

\section{Conclusion}

We have shown in this study how calcium hydroxide modifies significantly the mechanical behavior of a calcite paste.
Indeed, the initial elastic modulus of the suspension exhibits a minimum with the pH change resulting from the Ca(OH)$_2$ introduction, two orders of magnitude lower than the modulus of the pure calcite paste.
This non-monotonic behaviour stems from the competition between two opposite effects of the dissolved hydrated lime, that we have evidenced by zeta potential measurements and speciation calculations.
On the one hand, calcium ions complex at the calcite surface, increasing its surface charge and promoting thereby the inter-particle repulsion.
On the other hand, the added calcium hydroxide increases the ionic strength, leading to an increased screening of the surface charge, inhibiting inter-particle repulsion.
We have exemplified this competition in plotting the energy barrier of the interaction potential against pH, which exhibits a maximum at the exact pH where the elastic modulus is minimum.

In the long-term, the absorption of atmospheric CO$_2$ induces a carbonation of the lime that transforms into calcite, so that
all samples recover the mechanical properties of pure calcite in roughly 1 day.

We have also tested the influence of the addition of sodium hydroxide in the calcite suspension.
NaOH has been seen to inhibit the inter-particle repulsion, by both reducing the {absolute value of the} zeta potential, and increasing the ionic strength, which leads to a more rigid paste compared to pure calcite.

Overall, we found a direct correlation between the elastic modulus of colloidal calcite paste and the {energy} barrier calculated using the classical DLVO potential. This results contrasts with recent local force characterizations highlighting the non-DLVO nature of interaction forces between calcite surface in aqueous solutions \cite{royne_repulsive_2015, diao_molecular_2016, javadi_adhesive_2018, dziadkowiec_surface_2018}, attributed to to secondary hydration forces, ion-ion correlations or surface roughness. This could be due to the smoothness of our calcite colloids at the nanometric scale.

A practical consequence of our findings  is  that calcium hydroxide could be used as an admixture to get more workable suspensions of calcite colloids. The induced extra-fluidity spontaneously disappears by simple contact with air and the final mechanical properties are independent of the quantity of added lime. 
This tunability of the paste elasticity ---two orders of magnitude in elastic modulus--- is due to the nanometric scale of the particles and could allow the paste to be injected in porous minerals for consolidation, similarly to  nanolime binders that have been recently developed  for the restoration of limestone and lime-based historical heritage~\cite{daehne2013calcium}.

\section*{Acknowledgements}

This project has received funding from the European Union Horizon 2020 research and innovation program under the Marie Sk\l{}odowska-Curie grant agreement No.642976-Nano\-Heal Project. The results of this article reflect only the authors' view and the Commission is not responsible for any use that may be made of the information it contains.  CB acknowledges support from Institut Universitaire de France. 
{We thank Fernando Bresme and Juan Olarte Plata for sti\-mu\-la\-ting discussions, and Nicholas Blanchard for help with the TEM images.}

\bibliographystyle{model1-num-names}

\section*{References}

\bibliography{mybibfile}

\end{document}